# Title:

**Error catastrophe can explain the Fermi-Hart Paradox**

# Authors & Affiliation:

Axel Kowald (Axel.Kowald@ncl.ac.uk, Axel.Kowald@gmail.com)

Newcastle University, Institute for Ageing, Newcastle upon Tyne NE4 5PL, UK.

# To be published:

## Abstract

It has been argued that self-replicating robotic probes could spread to all stars of our galaxy within a timespan that is tiny on cosmological scales, even if they travel well below the speed of light. The apparent absence of such von Neumann probes in our own solar system then needs an explanation that holds for all possible extraterrestrial civilisations. Here I propose such a solution, which is based on a runaway error propagation that can occur in any self-replicating system with finite accuracy of its components. Under universally applicable assumptions (finite resources and finite lifespans) it follows that an optimal probe design always leads to an error catastrophe and breakdown of the probes. Thus, there might be many advanced civilizations in our galaxy, each surrounded by their own small sphere of self-replicating probes. But unless our own solar system has the extraordinary luck to be close enough to one of these civilizations, none of these probes will ever reach us.

Why is there no von Neumann probe on Ceres?

# Introduction

The so called Fermi Paradox is concerned with interstellar spaceflight and the existence of advanced extraterrestrial civilizations. If in the history of our galaxy intelligent extraterrestrial life-forms have developed with the ability of interstellar space flight, at least some of them would explore and colonise the galaxy. Even assuming moderate spaceship velocities (~0.01c), it would only take a few million years to reach and colonise the whole galaxy, including our own solar system. Since this seemingly has not happened, the questions is where are they?

Some authors argue that this means that we are the first and only civilization in the galaxy capable of space flight (Hart, 1975; Tipler, 1981), while others try to find less anthropocentric explanations. Among these ideas are Fermis doubts that interstellar space travel is feasible (Gray, 2015), the proposal that we live in a galactic zoo (Ball, 1973), that the whole universe as we see it is a simulation (Baxter, 2001), that economic constraints make space travel unlikely (Hosek, 2007), that the expansion rate is much slower than assumed (Haqq-Misra and Baum, 2009) or that colonies decide to refrain from space travel, leading to large uncolonized regions following a percolation process (Landis, 1998). Actually, Webb (2002) describes in his book 50 possible answers to the problem.

Although several of the ideas are good candidates to explain the absence of biological extraterrestrial colonists in our solar system, none of them can satisfactorily explain the absence of robotic extraterrestrial colonists. With this I mean the presence of a self-replicating von Neumann probe, as has first been proposed by Tipler (1981). Such a space probe would fly to a target star, search for an appropriate landing area and begin to produce several copies of itself, which then take off to visit the next solar systems. Of course it is well beyond our current capabilities to construct such a probe, but von Neumann (1966) showed its principle feasibility and Freitas (1980) estimated the size and composition of such a probe. Galactic exploration by such self-replicating robotic probes is quite different from a colonization process by biological lifeforms. Robotic probes have very low demands regarding a suitable place for replication, any small moon or asteroid will do. So while only few star systems might be suitable for biological colonisation, probes can virtually multiply in any system. Furthermore, they will never change their minds or question their mission. Once started they will never stop to follow their program. And the economic costs to construct such a probe will be much smaller than to build a spaceship for biological colonists. Indeed, as is discussed in the next section, the development of such probes might follow as a natural consequence of mining efforts in the solar system of the constructing species.

Probably the main aim of such a probe would be exploration and data gathering, thus it is doubtful if communication with us would be one of its mission targets. Freitas (1983) therefore correctly pointed out that it will be quite difficult to spot such a probe, should it already have arrived here in our solar system. However, that was more than 30 years ago and with the recent arrival of NASA's Dawn spacecraft at Ceres (www.nasa.gov/mission_pages/dawn/main), it is time to revisit this question. True, there are still vast areas

Why is there no von Neumann probe on Ceres?

of the solar system, which have not been visited by our spacecrafts, but the asteroid belt is probably one of the favourite landing places for a von Neumann probe. Even if we assume that such a probe has a propulsion system that is well beyond our current capabilities, it is reasonable to assume that planets with strong gravitation and disturbing atmosphere are being avoided. Thus, a moon or asteroid that is nevertheless large enough to provide all the building material for the replication process would be suitable. Since all the elements needed for the construction of daughter probes have to be obtained by mining, the required size can be substantial. There will also be an optimal distance from the sun. Too close and the temperature variations will cause unnecessary stress, too far away and the composition might be unsuitable (too few heavy elements). Considering these constraints a large object in the asteroid belt would seem to be an ideal candidate. Dawn has visited Vesta and Ceres, with 525km and 950km diameter, the two largest objects in the asteroid belt, but has so far not found any evidence for an extraterrestrial probe. The famous bright spots seen on Ceres would have been a prime candidate for a von Neumann probe, but unfortunately are most likely a mixture of ice and salt (Nathues et al., 2015).

So, it seems that there is indeed no self-replicating probe in our solar system and we are again faced with the question how we can explain this. In the remainder of this text I would like to suggest a new explanation that is based on the idea of an error propagation leading to an error catastrophe during successive rounds of replication. I will show that this is a problem, which all engineers of von Neumann probes are facing and thus error catastrophe might be a general explanation for the Fermi-Hart Paradox applied to robotic self-replicating probes.

## Error Propagation in Living Cells

Self-replicating probes are in many aspects very similar to living cells. Probes dissipate energy to keep themselves functioning and away from thermodynamic equilibrium, they disperse to a new habitat where they convert available resources (soil, minerals, chemical elements) into components (machines) of their own metabolism with the aim to assemble new copies of themselves. Like cells they have a master blueprint that specifies exactly how and when all the different components have to be fabricated. In cells that blueprint is the DNA, which encodes the information for all required proteins and metabolites, in the probe it is the mainframe computer that holds the construction plans for all the machinery. In view of this similarity it might be helpful to look at a serious obstacle that biological cells are facing during their process of self-replication.

In the 60ies and 70ies it was realised that cells might encounter a problem regarding a runaway error propagation inside their protein synthesis machinery (Orgel, 1963). The blueprints for all proteins are located on the DNA and when the cell needs to synthesise a new protein, a copy of its building instruction is created in form of a messenger RNA (mRNA). During the next step, large cell components, called ribosomes, bind to the mRNA and synthesize a new protein according to the information stored in the Why is there no von Neumann probe on Ceres?

mRNA (Figure 1a). Ribosomes, however, are themselves large complexes of proteins, which means that ribosomes are also responsible for synthesising new ribosomal proteins. I.e. ribosomes make copies of themselves (Figure 1b).

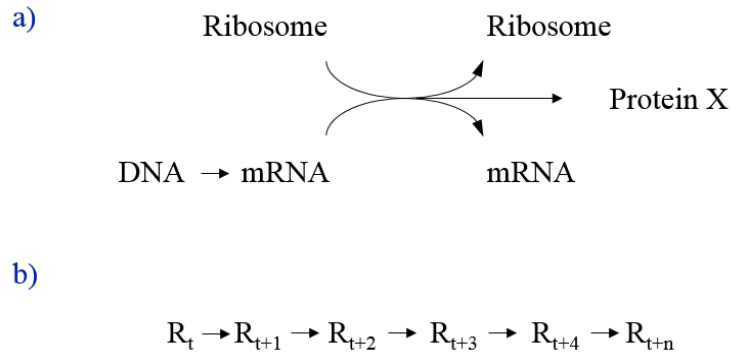

Figure 1: (a) Schematic representation of the protein biosynthesis in living cells. Ribosomes read the information that is stored on the messenger RNA (mRNA) to synthesise a new protein (Protein X). (b) Since ribosomes are also composed of proteins, ribosomes synthesise the next generation of ribosomes, which again build the next generation, and so on.

The problem is that ribosomes only work with a finite precision and sometimes introduce errors into the newly synthesised proteins. The question is then, what happens if ribosomes introduce errors in the next generation of ribosomal proteins? The mechanistic details of the protein biosynthesis mechanism are well understood, which made it possible to develop mathematical models to describe the resulting process of error propagation (Hoffman, 1974; Kirkwood and Holliday, 1975). The main result of these studies is that two outcomes are possible. If the ribosomal accuracy is below a critical threshold, the quality of the ribosomes deteriorates from generation to generation with the ribosomes becoming more and more error prone until protein biosynthesis collapses. This has been called "error catastrophe". But if the accuracy is above the critical threshold, the ribosomal error rate actually converges (from generation to generation) towards a stable steady state. It turns out that living cells normally operate above this threshold and thus do not suffer from an error catastrophe.

## Error Propagation in von Neumann probes

But this issue is also relevant for the replication process of von Neumann probes, since also here we have machines that make copies of themselves. In a cell there is a universal constructor (the ribosome) that produces all machines (proteins) that are part of the cell, while in a self-replicating space probe the situation is more complex, since often many heterogeneous parts have to be produced and assembled to form a working machine. However, at the end of a replication cycle a von Neumann probe generates an identical copy of itself and must thus face the consequences of error propagation. Also machines, like enzymes, only work with a finite accuracy leading to deviations during cutting, screwing, weighing, soldering, welding and so on. The question is how these deviations from the blueprint influence the performance of the probe of the next generation. Obviously, if the accuracy is too low not even the assembly of the first generation probe will succeed, if the accuracy is raised it might be possible to produce a first generation space probe that reaches its target star but then fails to reproduce itself. It is worth noting that this deterioration process is completely independent of any damage to the software that controls the space probe and the replication process. The error propagation we are discussing here, takes place without a single bit being changed in the space ships computer systems.

## A Toy Model

In case of the protein biosynthesis process inside cells, the system is known well enough to develop an adequate mathematical description to model the long term outcome of its error propagation. Unfortunately, our knowledge about the mechanistic details of self-replicating machines is rather limited, which makes it quite difficult to develop a mathematical model that can be called realistic. However, I nevertheless think that a toy model is helpful to illustrate the principle and basic consequences of an error catastrophe for the propagation of galactic colonisation via self-replicating probes. Let's assume that the functional status of the probe, S, declines from generation to generation according to the simple function,

$$S(n) = q^n$$

where "n" denotes the number of generations and "q" represents a quality parameter with q<1. We furthermore assume that the time, which is needed to replicate the probe, increases as S decreases. If S drops below a critical level, $S_d$, the replication process collapses completely and the production of new probes ceases. Effectively the probe is "dead". The following expression for generation time, T, formalises these assumptions, with "k" being a parameter controlling the overall replication speed.

$$T(n) = \frac{k}{S(n) - S_d}$$

Figure 2 shows that the functional status continuously declines from generation to generation, leading to a substantial increase of the generation time, T. For these simulations it was assumed that each replication results in a 5% loss of function (q=0.95), that probe assembly completely stops when S has dropped to 10% ($S_d$=0.1) and that the initial replication time is 500 years (similar to Freitas (1980)) resulting in k=425. With the chosen parameters the functional status reaches the critical breakdown level, $S_d$, after 45 generations, but the expansion rate of the colonisation process slows down right from generation one.

Why is there no von Neumann probe on Ceres?

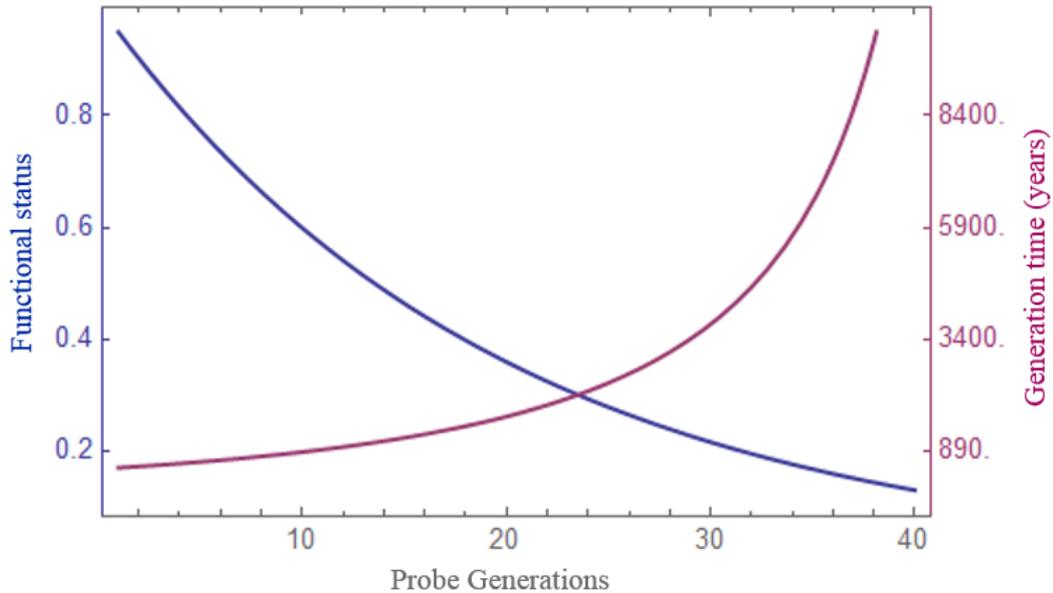

Figure 2: Diagram showing how functional status, S(n), and generation time, T(n), develop during 40 successive probe generations. The parameters used for this calculation were q=0.95, k=425 and $S_d$=0.1.

If we assume that the space probes visit all star systems around the originating civilisation and that the average distance between stars is 5 light years (LY), we can calculate how the radius of exploration grows with time. To calculate the time to reach a certain radius we calculate the sum over the individual replication times plus an inter-star travel time of 500 years (assuming a probe speed of 0.01 c).

$$t(r) = \sum_{n=1}^{r/5}(T(n)+500)$$

As can be seen from Figure 3 the colonisation process continuously decelerates. The first 100 LY are reached in 29000 years, but it takes another 114000 years for the next 100 LY. The expansion process asymptotically approaches a limit of 225 LY, which is never reached. That is the point when the functional status, S, has dropped to the threshold, $S_d$, which is insufficient to assemble a new space probe. Without error propagation a distance of 225 LY would have been reached after 45000 years (Figure 3, black line).

Why is there no von Neumann probe on Ceres?

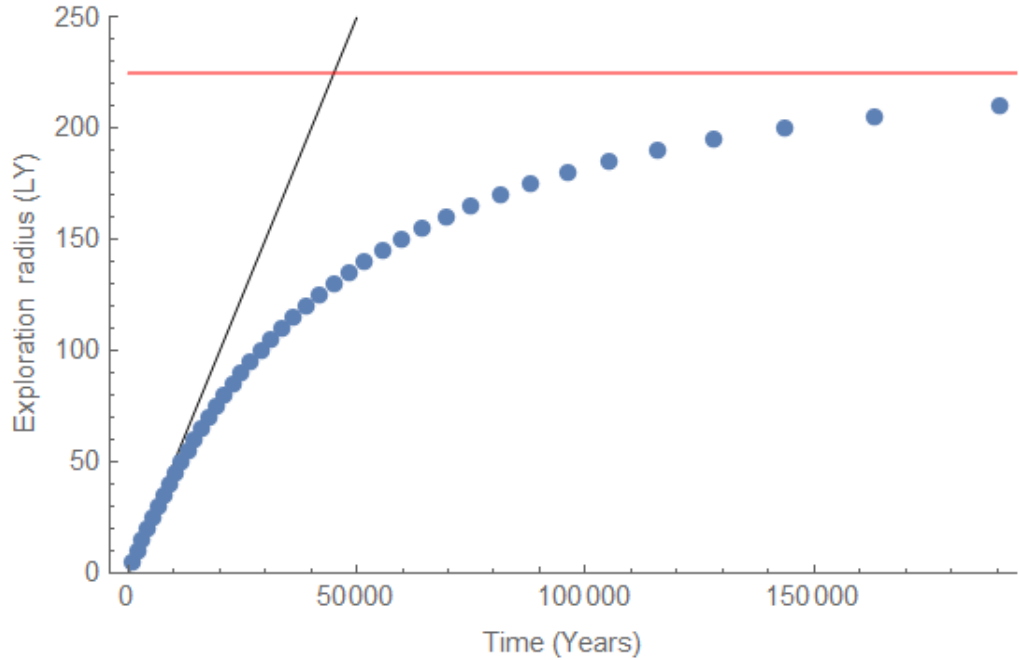

Figure 3: Diagram showing how far probes explore space around the originating star system with increasing time under the assumption that stars are 5 LY apart and the probe travels with 0.01 c. Other parameters as in Figure 2. The red line is the maximum distance that the colonisation process asymptotically approaches and the black line is the expansion rate without error catastrophe.

## Can an Error Catastrophe be avoided?

So far we have described that an error catastrophe can occur in self-replicating probes and that it can lead to a continuously deteriorating state of the fabricated probes, which finally brings the colonisation process to a halt. But is this not easy to avoid by adding quality checkpoints to the fabrication process? By simply increasing the precision during the production of all parts? And is the toy model not seriously flawed because it predicts for all allowed values of the quality parameter (q<1) a break down?

Let's make this clear. In principle it is of course possible to construct a probe that does not suffer from an error catastrophe, in the same way that living cells can reproduce without running into a catastrophic error propagation. However, I am going to argue that firstly there is actually no motivation whatsoever to build such a perfect probe and secondly that rational probe design will lead to a probe that operates in the low accuracy regime. Thus, the toy model is quite adequate since its purpose is to visualise the consequences of an ongoing error catastrophe inside self-replicating probes.

Why is there no von Neumann probe on Ceres?

## No need for a perfect probe

For the following discussion it is useful to briefly outline the economic justification for the development of a self-replicating probe. The point I like to emphasise is, that not everything that is technically feasible and scientifically interesting is also going to be funded. We only have to look at the time schedule of NASAs scientific missions to realise that many projects are delayed or cancelled because of budget constraints. Eventually such missions have to be financed through society and are in direct competition with other resource demanding projects, a point that has also been addressed by others (Hosek, 2007). However, it is not difficult to envisage a series of economically attractive steps, which end in the construction of a von Neumann probe.

Space exploration and exploitation, even in the local solar system, is a difficult and expensive undertaking. Thus, if machines could be used to help during the repair, fabrication or assembling of other machines or structures, it would lead to great savings in time and financial resources. The development of such an automation method would be a gradual process, possibly starting with remotely controlled machines to manufacture a moon base, via autonomous machines for mining projects on more distant asteroids and moons, to machines that are not only capable of mining minerals and ores, but that can also process these resources on-site to produce further machine parts. Such machines would be an ideal starting point for developing robotic probes to be send to neighbouring stars. Obviously those visits would be scientifically interesting, but such outposts could also be of military relevance as an early warning system. For both purposes it would be extremely useful if the probe would be capable to build new structures (e.g. forts, telescopes, power stations) on-site. This would reduce the machinery that has to be brought from the home system and it would be more flexible, because the probe can decide what is needed after a close inspection of the target system.

From here it is only a small step to turn such a probe into a fully self-replicating probe. If it can be instructed to build telescopes and power stations, it can also be instructed to build a copy of itself. All that is needed are the right blueprints in the probes computer. The logic is simple, for very little effort (creating the necessary blueprints), one can generate multiple generations of probes, which visit and investigate a growing sphere of stars around the home system. However, there is a diminishing return in the sense that the further the systems are away the less important is a successful visit by the probe. This is because they are less of a military threat, because the number of scientific discoveries decreases with the number of investigated stellar systems and most importantly, because the time a probe needs to reach further stars counts in the millennia. Assuming a travel speed of 0.01c and 500 years to build a new probe, it would take 40000 years to reach a star system 200 LY away without error catastrophe (Figure 3, black line).

Of course such time spans are negligible in comparison to the age of the galaxy, but to estimate the motivation of an extraterrestrial civilisation to undertake a project of such a duration, we have to compare the respective time span with the average life expectancy of individuals of the alien species. Hosek (2007) quite rightly points out that individuals with a finite lifespan tend to prefer benefits/projects/results that can be realised within their lifespan over benefits that might only be experienced by future generations. Unfortunately, we don't know the life expectancy of extraterrestrial lifeforms, but we can make an educated guess, since life histories of species everywhere in the galaxy are the result of an evolutionary process, quite similar to the one on earth. Lifespan is inversely related to the risk of the ecological niche in which a species lives and the maximum lifespan that is achieved by various long-lived animals (elephants, parrots, humans, whales, turtles) ranges roughly from 100-200 years (Finch, 1990). Furthermore, if we assume that humans (and advanced extraterrestrials) could completely abolish the aging process, average lifespan is boosted to roughly 1000 years (limited only by disease and accidents). If we thus take a few millennia as upper guess for the lifespan of intelligent extraterrestrials, we see that the inspection of more distant solar system via self-replicating probes quickly becomes less interesting because it is well beyond the individual time horizon.

Consequently, there will be a demand for self-replicating probes to explore neighbouring solar systems, but there is no demand for probes that are capable of truly indefinite reproduction.

## Rational probe design

In the last section I argued that there is no need to construct a perfect von Neumann probe. Here I'm going to explain why an optimised probe design automatically leads to a probe with a limited replication potential.

Because of the finite lifespan of biological lifeforms a main design priority of the probe will be to reduce the time required for travel and replication. Everything else being equal, there will be a replication accuracy, which minimises replication time and this accuracy will be below the critical threshold that avoids an error catastrophe. Figure 4 describes the situation for a first generation probe (i.e. without the effects of error propagation that manifests itself in later generations) in qualitative terms. If replication accuracy (i.e. the accuracy of the fabrication and assembling process of all the millions of components) is too low, not even a single functional probe can be produced and replication time will be infinite. If accuracy is increased, a new probe will be completed, but replication time is still quite long because the fabrication process will not run at full efficiency, since the components of the involved intermediate infrastructure, are manufactured with low accuracy. Now let's think about the other extreme, about a replication accuracy that is just high enough, so that the following generations of probes avoid an error catastrophe. As we have discussed, there is no demand for such a feature, so that the accuracy can be reduced a little bit. That means there will be fewer machine parts that are being rejected because of their insufficient quality/performance/accuracy. Overall, there might be fewer quality checkpoints (which themselves are implemented by machines that have to be build). And quite generally, if an output of lower quality is acceptable, such an output can normally be produced with machines of a simpler design. All this means that the reduction in accuracy leads to a smaller probe (i.e. the flying spacecraft plus all the intermediate technical infrastructure that is needed for building the spacecraft) and an increase in replication speed, i.e.

Why is there no von Neumann probe on Ceres?

to a reduction in replication time. From this follows that there will be a minimal replication time, which is associated with a replication accuracy that is in the regime leading to an error catastrophe.

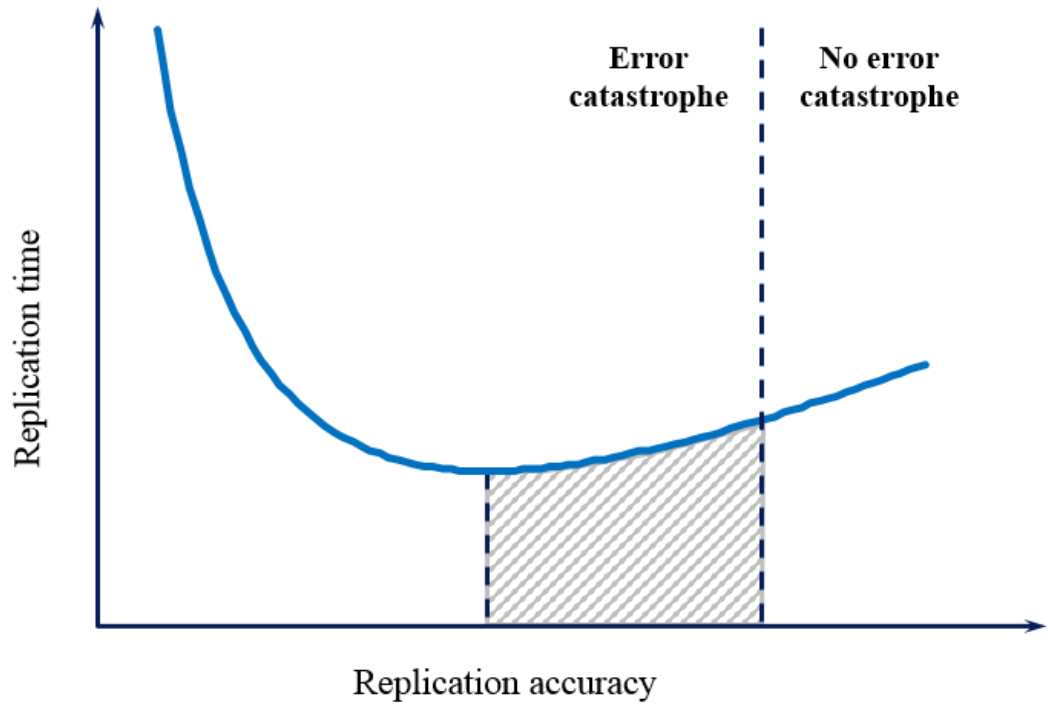

Figure 4: Relationship between a first generation replication time and replication accuracy of a von Neumann probe. If the accuracy is too low, replication cannot be completed at all and replication time will be infinite. But a too high accuracy also increases replication time because extensive quality control measures require time and additional hardware. A rational probe design leads to an accuracy within the shaded area. See text for details.

Now that we have rationalised the qualitative shape of the curve, we can ponder which replication accuracy will be chosen by the probe designers? The unknown engineers face a trade-off between replication time and severity of error propagation. The point of minimal replication time might not be suitable because it might lead to such a severe error propagation that already the next generation of probes is sterile in the sense that they cannot reproduce themselves. Increasing replication accuracy increases the size of the stellar neighbourhood that can be reached by the probes before an error catastrophe "kills" them, but it also increases replication time. Thus, depending on properties specific to the extraterrestrial civilisation (individual lifespan, fascination for space exploration, economic power, etc.), any accuracy within the shaded area of Figure 4 might be chosen, but in any case an optimal probe design always leads to probes that are prone to an error catastrophe!

Why is there no von Neumann probe on Ceres?

## Summary and Discussion

The spread of self-replicating robotic probes in the galaxy is a particularly difficult version of the Fermi-Hart paradox. Particularly difficult because robotic probes can multiply in any solar system and follow a strict program. Hence, the apparent absence of such von Neumann probes in our own solar system is especially difficult to explain. Here I proposed a solution for this problem, which is based on a runaway error propagation that can occur in any self-replicating system with finite accuracy of its components. The phenomenon has been recognised and studied in living cells and it could be shown that this process either leads to an error catastrophe and breakdown, or that the system approaches a stable error level with an infinite replication potential.

The finite lifespan of any extraterrestrial lifeform automatically leads to a preference of results in the near future over those in the far future and such a preference also arises from simple economic constraints. That means to judge the feasibility of large scale projects, it is important to compare the project timescale to the lifespan of the engineers and not to the age of stars or the galaxy. Then it becomes clear that an infinite replication potential is not a design priority for a hypothetical self-replicating probe. Instead, a short replication time will be a design priority and it turns out that there is a trade-off between replication time and the severity of the error propagation process and hence the volume of space that can be explored by the probes. However, for any reasonable estimate of the lifespan of the probe engineers, the optimal replication accuracy of the probe will always be in a range that finally ends in an error catastrophe (Figure 4).

The arguments that lead to these conclusions are based on two assumptions that hold for all extraterrestrial civilisations at any place and at any time. The first assumption is that biological lifeforms have a finite life expectancy. I made in the previous sections a generous educated guess for an upper limit, but even if that guess would be off by a large factor, it would not change the basic point. The second assumption is that there is always a finite amount of resources that can be assigned to different projects and that there are, consequently, constraints arising from optimising the allocation of these resources (see also Hosek (2007)). Even if mankind had 10 times more energy, computing power or personnel, it would still make sense to finish a project in half the time, if possible. Of course, the explanation given here does not make the construction of a probe with unlimited replication potential impossible, but it provides a rather profane rational explanation why no civilisation is ever motivated to do so.

These assumptions imply that von Neumann probes are constructed with limited replication capabilities, which are sufficient for the needs of the constructing civilization. A simple mathematical model shows that, as a consequence, the expansion rate of the probes continuously slows down until it comes to a halt. Thus there might be many advanced civilizations in our galaxy, each surrounded by their own small sphere of self-replicating probes. But unless our solar system is by chance close enough to one of these civilizations, we will never see an extraterrestrial self-replicating probe, which is also the reason why there is no von Neumann probe on Ceres.

Why is there no von Neumann probe on Ceres?

# Acknowledgements

The author would like to thank T. Richter and TBL. Kirkwood for interesting and helpful discussions.